\def\CII{\mbox{C\,{\sc ii}}} 
\def\CIV{\mbox{C\,{\sc iv}}} 
\def\NV{\mbox{N\,{\sc v}}} 
\def\OI{\mbox{O\,{\sc i}}} 
\def\SiII{\mbox{Si\,{\sc ii}}} 
\def\SiIV{\mbox{Si\,{\sc iv}}} 
\def\keV{\mbox{ke\kern-.1emV}} 

\documentclass{aa} 
\usepackage{graphicx} 
\begin{document} 

% \thesaurus{03(11.06.1, 12.07.1, 12.03.3)} 

\title{Gravitationally lensed high redshift galaxies 
in the field of 1E0657-56 
\thanks{Based on observations obtained with 
FORS at Paranal at UT1 of the VLT.}} 

\author{D.\,Mehlert\inst{1} 
\and S.\,Seitz\inst{2} 
\and R.\,P.\,Saglia\inst{2,3} 
\and I.\,Appenzeller\inst{1} 
\and R.\,Bender\inst{2} 
\and K.\,J.\,Fricke\inst{4} 
\and T.\,L.\,Hoffmann\inst{2} 
\and U.\,Hopp\inst{2} 
\and R.-P.\,Kudritzki\inst{2,5} 
\and A.\,W.\,A.\,Pauldrach\inst{2}} 

\offprints{D.\,Mehlert,\\
\email{dmehlert@lsw.uni-heidelberg.de}} 

\institute{Landessternwarte Heidelberg, K\"onigstuhl, 
  D-69117 Heidelberg, Germany 
\and Universit\"atssternwarte M\"unchen, Scheinerstra{\ss}e~1, 
  D-81679 M\"unchen, Germany 
\and Research School of Astronomy and Astrophysics, 
  The Australian National University, Cotter Road, 
  Weston Creek, ACT 2611, Australia 
\and Universit\"ats-Sternwarte G\"ottingen, Geismarlandstra{\ss}e~11, 
  D-37083 G\"ottingen, Germany 
\and University of Hawaii, Institute for Astronomy, 2680 Woodlawn Drive, 
  Honolulu, Hawaii 96822, USA} 

\date{accepted on September, 14th 2001} 

\authorrunning{Mehlert et al.} 
% \titlerunning{High redshift galaxies} 

\abstract{We present images and long-slit spectra obtained with FORS1 
at UT1 of the VLT centered 
%%%RPS avoid the repetition of ``images'' 
on the gravitational arc 
of the galaxy cluster 1E0657-56 ($z = 0.296$).  The cluster is one of 
the hottest, most massive clusters known so far and 
%%%RPS nothing is perfect! 
acts as a powerful gravitational telescope, amplifying the flux of 
background 
sources by up to a factor of 17.  We present photometric results 
together with the spectra of the gravitational arc ($z = 3.24$) 
and four additional amplified high redshift objects ($z = 2.34$ 
to 3.08) that were also included in the slit by chance coincidence. 
A magnification map has been obtained from a lens model derived from 
the multiple image systems.  We compare our observed spectra with 
models and briefly discuss the stellar contents of these galaxies. 
Furthermore we measured the equivalent widths of the \CIV~1550 and 
\SiIV~1400 absorption lines for the objects behind 1E0657-56 
%%%RPS not all of them... 
studied here, as well 
as for some additional starburst galaxies (nearby and at high $z$). 
For \CIV\ we find an increasing absorption equivalent width with 
decreasing redshift.  We discuss whether this correlation could be 
related to the increase of metallicity with the age of the universe.} 

\maketitle 
% \markboth{Mehlert et al.: High redshift galaxies}{} 


\section{Introduction} 
\label{introduction} 

In the recent years the technique pioneered by Steidel and Hamilton 
(\cite{SH92}, \cite{SH93})  of deep imaging to identify U- 
and B-drop-out objects has been extremely successful in finding 
high-redshift galaxies.  More than 700 galaxies with redshifts 
higher than 3 are known to date (Giavalisco et al.~\cite{Getal98}). 
Parallel to these studies, an increasing number of high redshift 
objects has been found in the background of massive clusters 
of galaxies.  The clusters act as a ``gravitational telescope'' 
(Pell\'o et al.~\cite{PKBFMBCDM01}), magnifying and stretching the 
background galaxies.  In particular, objects near the critical lines 
can benefit from large amplification factors (1 to 3 magnitudes) 
which ease spectroscopic follow-up observations.  In addition, 
sometimes the gain in angular resolution allows studying the internal 
structure and kinematics of these usually rather compact galaxies. 
Spectacular examples of such studies are the pair of $z = 4.92$ 
objects in the field of CL1358+62 (Franx et al.~\cite{Franx97}), 
the gravitational fold arc galaxy cB~58 (Seitz et al.~1998, 
Carlberg et al.~1996) which also shows the features of a classical 
Lyman break object (Pettini et al.~2000), the arcs in A2390 (Frye 
\& Broadhurst~\cite{FB98}, Pell\'o et al.~\cite{PKBBETBMSSB99}), 
the star-forming galaxy in A2218 (Ebbels et al.~\cite{E96}.) Here 
we report on five high-redshift, gravitationally lensed objects we 
have found in the background of the cluster 1E0657-56. 

Tucker et al.~(1998) identified the extended Einstein source 1E0657-56 
with a cluster of galaxies at a redshift of $z = 0.296$ showing a 
velocity dispersion, measured from 13 galaxies, of $\sigma_{\rm gal} 
\approx 1200\,{\rm km/s}$.  X-ray data from ROSAT and ASCA indicate 
that merging of at least two subclusters occurs in this highly luminous 
X-ray cluster.  Tucker et al.~(1998) derive a temperature of the hot 
gas in \mbox{1E0657-56} of $kT \approx 17\,\keV$.  Note, however, 
that Yaqoob (1998) reanalyzed the X-ray data and derived a temperature 
of $kT \approx 12\,\keV$ only.  This still makes the cluster one of 
the hottest known yet.  The total mass of 1E0657-56 within 1~Mpc, 
derived from the X-ray data, is about $2 \times 10^{15} M_{\sun}$. 
Optical images of 1E0657-56 show the presence of a large gravitational 
arc approximately $1'$~NW of the cluster's center as well as several 
fainter arcs at other locations.  In the following we describe the 
results of a photometric and spectroscopic study of the prominent arc 
(hereafter referred to as ``the arc'') and some neighboring objects. 
% The following three sentences could well be left out... 
In Sect.~\ref{observations} we present the observations 
and the data reduction. 
In Sect.~\ref{results} we discuss the results. 
In Sect.~\ref{conclusions} we draw our conclusions. 

\section{Observations and Data Reduction} 
\label{observations} 

\begin{figure*} 
\includegraphics[width=13cm]{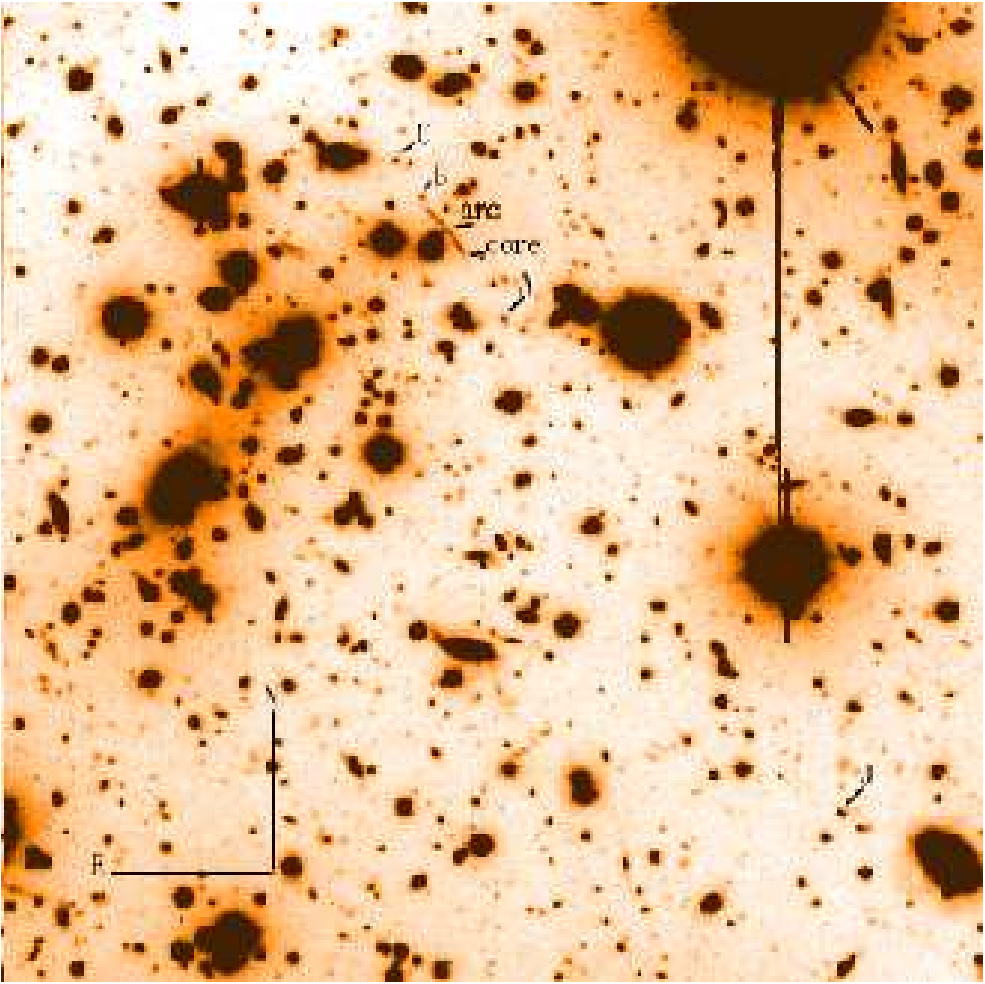} 
\hfill\parbox[b]{4.5cm}{ 
\caption{ 
The central $3.2' \times 3.2'$ of one 500~sec R-band image of 
the galaxy cluster 1E0657-56, taken with seeing of $0.75''$. 
The position of the arc, of its core, and of the objects A, B, C, 
and J (see Sect.~\ref{observations}) are indicated.  North at the top, 
East to the left. 
\label{fig:cluster} 
}} 
\end{figure*} 

The cluster 1E0657-56 was observed as a test object during the 
commissioning of the Focal Reducer Spectrograph (FORS1) at the Very 
Large Telescope (VLT) in December~1998 
%%%RPS 1999-->2000 
(Appenzeller et al.~2000). 
Images in the B (1500~sec), g (600~sec), R (3800~sec), and I 
(1200~sec) filters were collected using the standard resolution mode 
(0.2~arcsec/pixel), with seeing of 0.6--0.9 arcsec.  During the same 
run long-slit spectra of the prominent arc (see Fig.~\ref{fig:cluster}) 
were acquired with FORS1, using the grism I150 and a slit width 
of $1''$.  We covered the spectral range from $\lambda = 3300$ 
to 9200\,\AA\ with a spectral scale of 5\,\AA/pixel.  In total we 
gathered 3.5\,h of exposure time with an effective seeing of about 
$0.75''$.  Standard reduction (bias subtraction, correction for flat 
field variation, cosmic ray elimination) for both the photometry and 
spectroscopy data, as well as rebinning to the observed wavelength 
for the spectra, was performed using MIDAS routines.  The B, g, R, and 
I images were calibrated using photometric standard sources observed 
during the commissioning.  The accuracy of the relative calibration 
could be increased considerably by deriving the color-color diagrams 
% 
%%%RPS What is this library of stars? ref? 
% 
of the library of stars 
in the  FORS-filter system and comparing them 
to the numerous unsaturated stars observed in our field.  For the 
spectroscopy, spectrophotometric standards were observed to flux 
calibrate the spectra. 

Figure~\ref{fig:cluster} shows the central $3.2'\times3.2'$ of one of 
% habe hier die klammer eingefuegt, 5.juli 2001, von stella 
the R-band images (with a depth of 500~sec exposure time).   
Six objects are marked: the gravitational arc, 
visible to the northeast of the image center, with the bright knot 
(``core'') at its southwest end, objects B and C to the northeast 
of the arc, objects A and J to the southwest.  The magnitudes of the 
objects A, B, C, and J have been obtained using the SExtractor code 
(Bertin \& Arnouts~\cite{BA96}). 
%% hier habe ich noch den satz mit den reference daten rein, 5.juli 
%% 2001, stella 
Since the R-band data are by far the 
deepest we use them for the detection of the sources mentioned above 
and for the definition of the surface-brightness-contours of the arc. 
The `automatic' magnitudes in the 
R-band are shown in Column~2 of Table~\ref{tab:mag}, and those derived 
within an aperture of 3 arcseconds are given in Column~5. The data in 
the B, g, and I-band were convolved to the R-band seeing 
%%%RPS 0.8-->0.82 
($0.82''$) and 
the magnitudes within $3''$-apertures are shown for the convolved data 
%%%RPS  added ``respectively'' for clarity 
in the Columns 3, 4, and 6, respectively.

\begin{figure} 
\includegraphics[width=\columnwidth]{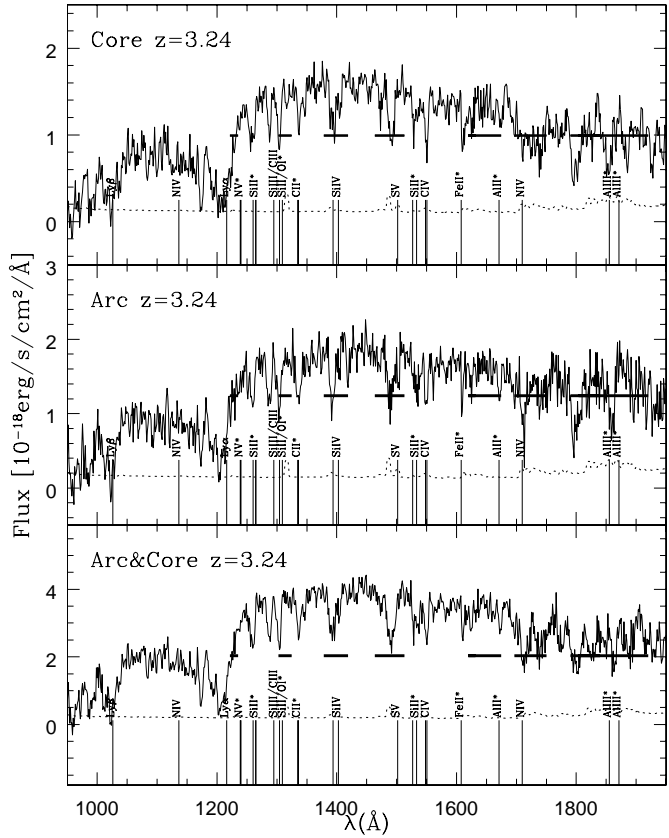} 
\caption{ 
Flux-calibrated spectra of the arc ($10''$, upper panel) and the core 
($3''$, lower panel).  The Ly$\alpha$ absorption line as well as the 
position of some expected metal absorption lines are indicated. Lines 
mainly caused by the interstellar medium are indicated by an asterisk. 
The horizontal bars indicate the position of prominent night sky 
blends (blueshifted to the rest frame of the galaxy).  The dotted 
lines indicate the $1\sigma$ (per pixel) error level of the flux. 
\label{fig:spec_arc} 
} 
\end{figure}

\begin{figure} 
\includegraphics[width=\columnwidth]{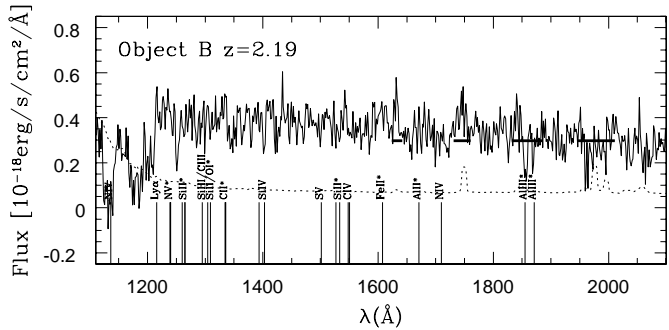} 
\caption{ 
Spectrum of Object~B, which lies a few arcsec NE of the arc. 
Ly$\alpha$~absorption and metal lines are indicated as in 
Figure~\ref{fig:spec_arc}.  The horizontal bars and dotted lines have 
the same meaning as in Figure~\ref{fig:spec_arc}. 
\label{fig:spec_b} 
} 
\end{figure}

\begin{figure} 
\includegraphics[width=\columnwidth]{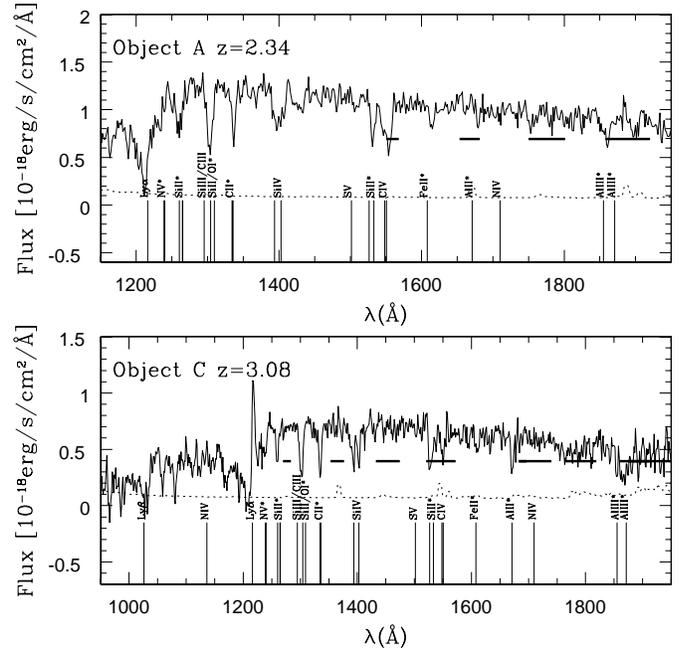} 
\caption{ 
Spectra of objects A (upper panel) and C (lower panel), which lie 
about $20''$~SW and $10''$~NE of the arc, respectively. 
Ly$\alpha$~absorption and metal lines are indicated as in 
Figure~\ref{fig:spec_arc}. 
\label{fig:spec_ac} 
} 
\end{figure}

\begin{figure} 
\includegraphics[width=\columnwidth]{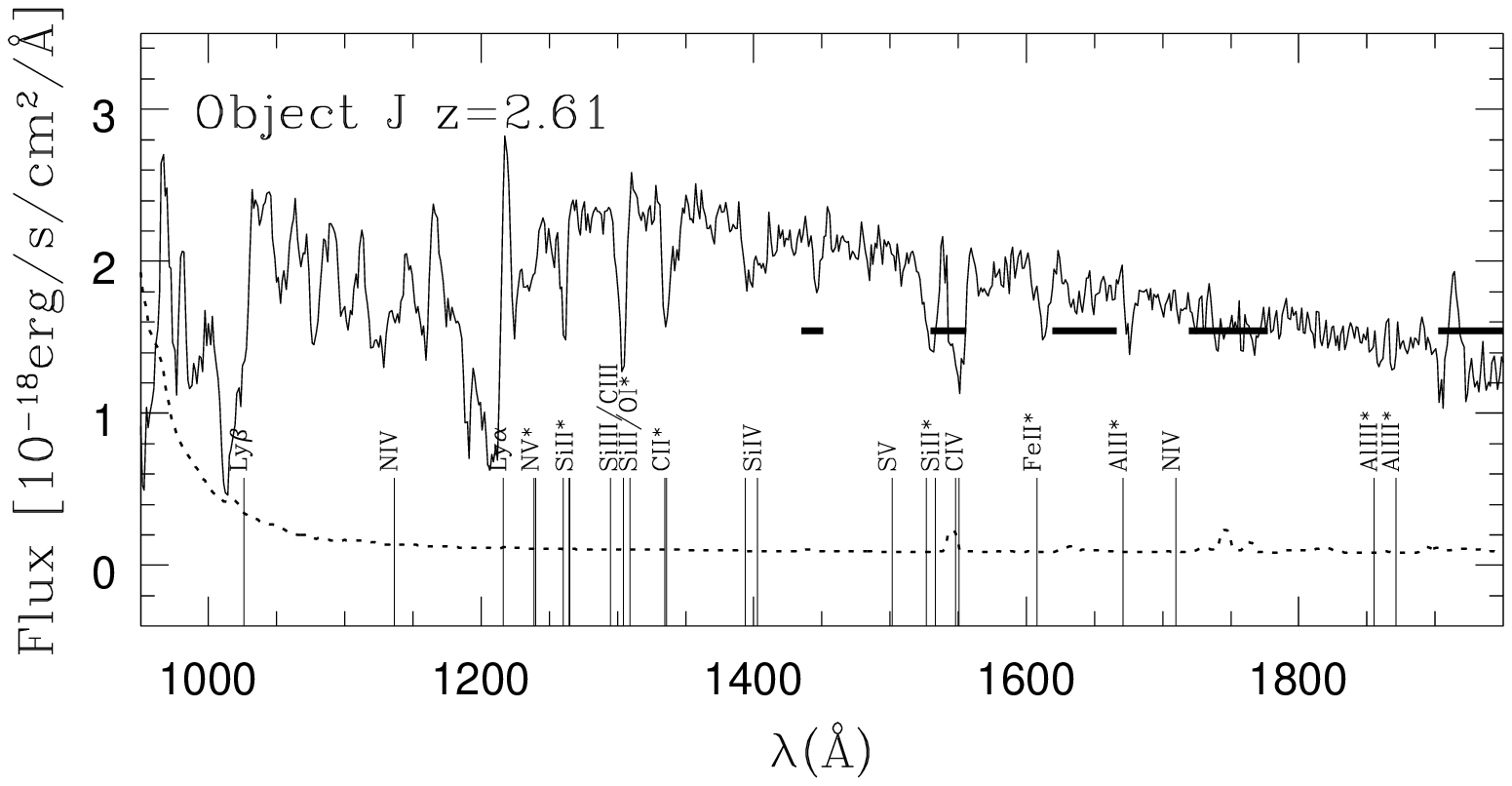} 
\caption{ 
Spectrum of object~J, which lies approximately $2.5'$~SW of the arc. 
Ly$\alpha$~absorption and metal lines are indicated as in 
Figure~\ref{fig:spec_arc}. 
\label{fig:spec_j} 
} 
\end{figure}

%stella/ralf bitte lest diesen Absatz nochmal genaz genau und checkec ob das 
%so stimmt. Gegebenenfalls bitte ergaenzen oder aendern. 
%Ralf - kannst du nochmal die jeweiligen zphot fuer arc und core fuer die drei 
%verschiedenen Isophoten in Tabelle 1 eintragen. Das war wohl einer der 
%Hauptanliegen des referees bei seiner Frage nach den Fehler. 
%Frage: wenn die phot z fuer core und arc zwischen 2.7 und 3.4 variieren, 
%warum nehmen wir dann 3.4 und nicht nen Mittelwert? Gibt es dafuer nen 
%speziellen Grund? 
% 
% 
% doerte, das mit den 2.7 und 3.4 stimmt so meiner meinung nach nicht 
% ralf war da zu konservativ. ich habe alle quellen angeschaut, die 
% die farben haben wie die 3 arc-measurments. das ergebnis 
% entweder werden sie als low z mit z<0.5 klassifiziert, oder sie 
% haben z=3.0-3.3 !! 
% 
% der eintrag unten kommt von mir, stella, 5.juli 2001 
% ich bin dafuer, in der tabelle bei den photz des arcs 3.0-3.3 
% reinzuschreiben!! 
% 
Photometric redshifts were derived 
using the method presented in Bender et al. (2001). 
This method is a template matching algorithm rooted 
%%%RPS Baysian --> Bayesian 
in Bayesian statistics and resembles the method presented by Benitez 
(2000). The templates are semi-empirical and derived from the observed 
spectral energy distributions of galaxies in the Hubble Deep Fields. 
Except for the arc and the core we used the 3''-aperture colors to 
determine 
the photometric redshifts of the observed objects. For the arc and core we 
used the magnitudes of within $\mu_{\rm R}=25.0$, 25.5, 
and 26 magnitudes per square arcsecond, which corresponds to 8.3, 
5.2, and 3.2-sigma contours above the sky background. 
The photometric redshifts for the different isocontours of the arc and 
core 
% 
% ab hier habe ich nochmal editiert 
% 
varied between 3.0 and 3.3 (see Table~\ref{tab:mag}). 
These differences are smaller than the   
typical error of the photometric redshift for objects at 
$z > 2.5$, which is about 0.3 for the filters we have used here. 
Hence the differences in the colors for the arc measured within the 
3 different isodensity-contours are irrelevant for the 
photometric redshifts derived. 
% We have considered all objects which 
%have B-g, R-I and B-I-colors in the error-interval defined by the 
%3 `arc-colors'. The redshifts of these objects are all in within 
%3.0-3.3, or at low redshift, $z<= 0.4-0.5$. 
% 
Figures~\ref{fig:spec_arc} to~\ref{fig:spec_j} show the flux-calibrated 
spectra of 5 high-$z$ objects extracted from the longslit. The 
signal-to-noise ratios vary from 3 (Object~B) to $\ga 10$ (Object~J) 
per pixel, or about 14 to 50 per spectral resolution element.  The 
most prominent feature visible in all the spectra is the Ly$\alpha$ 
absorption line, but we also find several metal absorption lines. 
These lines -- especially those from the interstellar medium indicated 
in Figs.~\ref{fig:spec_arc} to \ref{fig:spec_j} -- establish the 
redshifts of all objects to be larger than~2.  Furthermore, we see 
that the photometric redshifts (see Table~\ref{tab:mag}) are in good 
agreement with those derived from the spectra. 

\begin{table*} 
\caption{ 
Photometry of the high-redshift galaxies. 
Column~1: object name (see Fig.~\ref{fig:cluster}). 
Column~2: The SExtractor automatic magnitudes in the R~band. 
Columns~3--6, 
  lines 1--6: The magnitudes of the arc and core 
  in the B, g, R, and I~band data are given within the 
  $\mu_{R} = 25.0$, $25.5$, and $26.0~{\rm mag/square ''}$ 
  surface brightness isophotes; 
lines 7--10: The cB, cg, R, and cI magnitudes within an 
  aperture of $3''$ diameter; the symbol ``c'' indicates that 
  the B, g, and I~band images have been convolved to the 
  seeing of the R~band image for these measurements. 
Column~7: gravitational amplification of the 
          background source in magnitudes. 
Column~8: photometric redshift obtained using the 
          colors derived  from Columns~3--6. 
\label{tab:mag} 
} 
\begin{center} 
\begin{tabular}{|r|c|c|c|c|c|r|c|} 
\hline 
Object &   &$\rm B_{iso}$ & $\rm g_{iso}$ & $\rm R_{iso}$ & 
$\rm I_{iso}$ & Magn. & $z_{\rm phot}$\\ 
&   &(mag) & (mag) & (mag) & (mag) &  (mag) & \\ 
\hline 
$\rm Arc_{\mu_R\le 25.0}$   &       & $24.37\pm 0.04$ & $23.49\pm 0.03$ 
& $22.20\pm 0.01$ & $21.74\pm 0.02$ &  & 3.3   \\ 
$\rm Arc_{\mu_R\le 25.5}$   &       & $24.16\pm 0.04$ & $23.36\pm 0.04$ 
& $21.93\pm 0.01$ & $21.49\pm 0.02$ &   & 3.2 \\ 
$\rm Arc_{\mu_R\le 26.0}$   &       & $24.06\pm 0.03$ & $23.31\pm 0.03$ 
& $21.81\pm 0.01$ & $21.39\pm 0.01$ &   & 3.1 \\ 
$\rm Core_{\mu_R\le 25.0}$  &       & $23.98\pm 0.03$ & $23.10\pm 0.03$ 
& $21.79\pm 0.01$ & $21.32\pm 0.01$ & $\approx 3.3$  & 3.3 \\ 
$\rm Core_{\mu_R\le 25.5}$  &       & $23.79\pm 0.03$ & $23.00\pm 0.03$ 
& $21.55\pm 0.01$ & $21.11\pm 0.02$ & $\approx 3.3$  & 3.2 \\ 
$\rm Core_{\mu_R\le 26.0}$  &       & $24.71\pm 0.02$ & $22.94\pm 0.02$ 
& $21.43\pm 0.01$ & $21.01\pm 0.01$ & $\approx 3.3$   & 3.1 \\ 
\hline 
Object & R$_{\rm aut}$ &cB ($3'')$ & cg ($3'')$ & R ($3'')$ & cI ($3'')$ 
& Magn. & $z_{\rm phot}$\\ 
& (mag) &(mag) & (mag) & (mag) & (mag) &  (mag) & \\ 
\hline 
A     &  $22.55\pm 0.01$ & $23.47\pm 0.01$ & $23.11\pm 0.01$ & 
$22.66\pm 0.01$ & $22.45\pm 0.01$ & 1.9--2.2 & 2.4 \\ 
B     &  $23.99\pm 0.02$ & $24.55\pm 0.04$ & $24.43\pm 0.03$ & 
$23.84\pm 0.02$ & $23.81\pm 0.05$ & 2.5--2.8  & 2.4 \\ 
C     &  $22.71\pm 0.01$ & $24.75\pm 0.04$ & $23.70\pm 0.02$ & 
$22.83\pm 0.01$ & $22.44\pm 0.02$ & 2.3--2.6  & 3.2 \\ 
J     &  $22.04\pm 0.01$ & $22.98\pm 0.01$ & $22.48\pm 0.01$ & 
$22.16\pm 0.01$ & $22.00\pm 0.01$ & 0.95--1.0  & 2.6 \\ 
\hline 
\end{tabular} 
\end{center} 
\end{table*}

%\begin{figure*} 
%\includegraphics[width=\columnwidth]{figure2.ps} 
%\caption{ 
%The Flux calibrated spectra of the Arc, its Core, Object A, B, C, and J 
%(see Fig. 1). The positions of the Lyman and of some metal absorption 
%lines are shown redshifted as dotted vertical lines.  The horizontal 
%bars indicate the position of prominent night sky blends The dotted 
%line shown for Object J indicates the size of the $1\sigma$ error level 
%of the flux. {\it We should overplot the magnitudes in B, g, R, I.} 
%\label{fig:spec} 
%} 
%\end{figure*} 

\section{Results and Discussion} 
\label{results} 

\subsection{Photometry} 
\label{results_phot} 

\begin{figure*} 
\includegraphics[width=13cm]{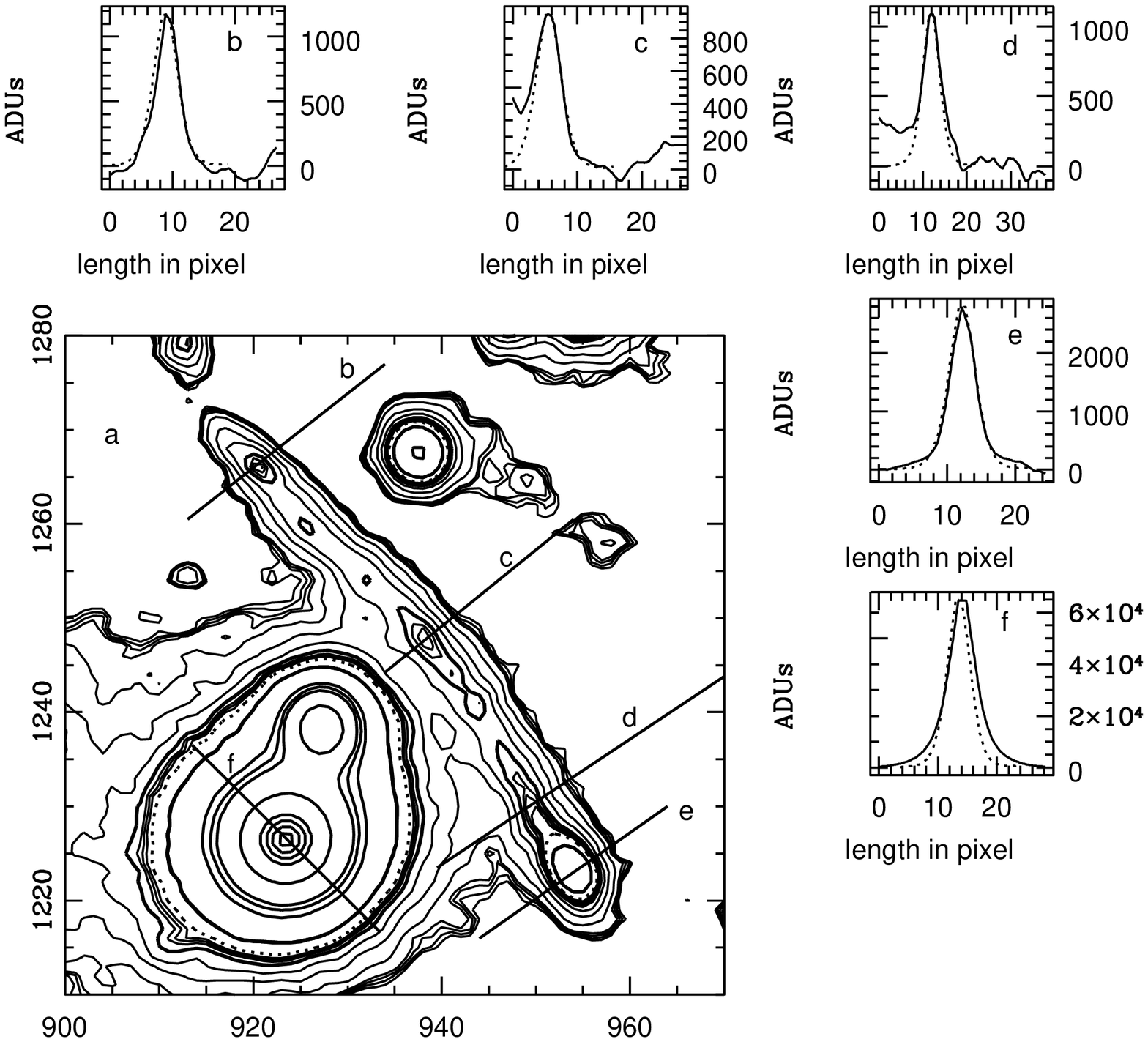} 
\hfill\parbox[b]{4.5cm}{ 
\caption{ 
{\bf (a)} Surface brightness contours of the 
arc in the R band (without background subtraction). 
We can identify three high-surface-brightness spots (b, c, d), 
which arise from the same part of the source. 
The compact and high-surface-brightness component at the SW (e) 
is single-imaged 
%%%RPS added ``and'' 
and called the ``core'' in the following. 
{\bf ~(b, c, d)} The brightness profile (after background subtraction) 
through the three spots of the arc indicated by the cuts. 
The comparison with a star (scaled to the same maximum, dotted curves) 
demonstrates that the arc is unresolved in its width. 
{\bf ~(e)} The brightness profile of the core. 
It is unresolved and stretched along the direction of the arc; 
the major-to-minor axis ratio of the dotted 
contour shown is about 1.55. 
{\bf ~(f)} The profile of the bright galaxy (SE of the arc), 
which clearly has a non-stellar brightness profile. 
\label{fig:arc} 
}} 
\end{figure*}

Inspection of Fig.~\ref{fig:cluster} shows that the arc is located 
NE of the image center (NW of the cluster's center).  The arc's 
brightness profile perpendicular to its major-axis is stellar, i.e., 
its FWHM agrees with that of a star (the FWHM for the data is $0.82''$ 
for the R-band,  $0.80''$ in the B-band,  $0.68''$ for the g-band and 
$0.6''$ for the I-band. 
%% eintrag oben von stella, 5.juli2001 
%% Bild 1 stellt nur einen Teil der Daten dar, wo die seeing qualitaet 
%% eben besser war naemlich 0.75'', die Tiefe der Daten betraegt nur 
%% in dieser Figur nur 500sec. 
%% 
% bitte lasse die 0.75'' stehen! 
%stella - hier die genauen seeing-Werte aller Filter. In der Caption von 
%Bild 1 hatten wir bislang fuers R Image ein seeing von 0.75 angegeben. 
%Weisst du wo der her kommt? Ich habe den Wert jetzt in 0.82 geaendert - o.k? 
%Ausserdem will der referee, das wir ``... briefly discuss on the choice of 
%the reference image.'' Irgedwelche anderen Kriterien ausser dem seeing. 
%Sollte man vielleicht kurz anmerken, damit er zufrieden ist 
). 
The two-dimensional 
R-band flux distribution of the arc is shown in more detail in 
Fig.~\ref{fig:arc}. Within the lowest contours shown, its length and 
width are $14''$ and $1.8''$, respectively. 
% Innerhalb der schwarzen Kontur ist Laenge $12.3''$ und 
% Breite $0.55''$ 
In the following we distinguish a pure ``arc'' and the ``core'' at 
its SW end. The arc shows substructure, in particular three `spots' 
with approximately the same surface brightness (after background 
subtraction), which we interpret to arise from the same source region. 
The brightness profiles through these spots along the indicated 
cuts in Fig.~\ref{fig:arc}a are shown (after background subtraction) 
in sub-figures b, c, and d of Fig.~\ref{fig:arc} (solid lines). The 
brightness profile of a star has been scaled to the same maximum value 
and added in each case (dotted line). The comparison demonstrates 
that the arc is unresolved along these cuts. Fig.~\ref{fig:arc} shows 
the profile of the core component that is unresolved in width as well 
with a major to minor axis ratio of 1.55 within the dotted contour. 
For more illustration we have added a similar sub-figure (f) showing 
the brightness profile of the bright galaxy in the Fig.~\ref{fig:arc}a, 
which clearly is non-stellar. We conclude from the images that 
the lensing configuration is the following: the core of the lensed 
galaxy is near but outside the cusp-caustic, and thus is mapped in 
only one image called ``core'', whereas the outer region touches the 
cusp-caustic and is mapped into three merging images (for comparison, 
see the cusp-arc in A370; Fort \& Mellier~\cite{FM94}).  Hence, in 
the 1E0657-56 arc, we see a high resolution map of the outskirts of 
a $z\approx 3$ galaxy. 

The comparison of the spectra of the arc and the core (see 
Fig.~\ref{fig:spec_arc}) shows that the two objects indeed have the 
same background source.  In contrast, object~B (which is located at 
the NE end of the arc) has a different redshift (Fig.~\ref{fig:spec_b}) 
and other colors (Table~\ref{tab:mag}) and hence does not belong to the 
same background source as the core/arc system. This fact additionally 
constrains the lensing model for 1E0657-56. 

\subsection{Lensing model} 
\label{lensmodel}

Using a simple lens model for the cluster, we derived a magnification 
of the original background sources {\bf by a factor up to 20 in flux.} 
A detailed lensing model will be discussed in a separate paper (Seitz 
et al.~2001).  Briefly, all mass distributions have been modelled 
either by a Singular Isothermal Sphere (SIS, with velocity dispersion 
as free parameter) or with the more general Elliptical Isothermal 
Sphere with a Core radius as described in Seitz et 
al.~(\cite{SSBHBZ98}) (EISC, with velocity dispersion, ellipticity, 
orientation and core radius as free parameters). The best-fitting 
model parameters have been found using the arc and multiple images in 
the same way as described in Seitz et al.~(1998). 
Multiple  images have been selected by there optical appearance (arc-like,
`broken arc', parity reversed images), and cross checking
surface-brightness and colors. Since the multiple images are extended,
we have identified special positions (like knots in the arc, `head' or
`tail' of an image), within
every image belonging to a image-configuration, and defined these
positions as `the' positions for the lens modelling. This explains why
the defined positions of the multiple images are not necessarily in agreement
with the center of light of these objects.

%stella - da stehts ja mit dem Arc - sieh nochmal unten 

\begin{figure*} 
\includegraphics[width=13cm]{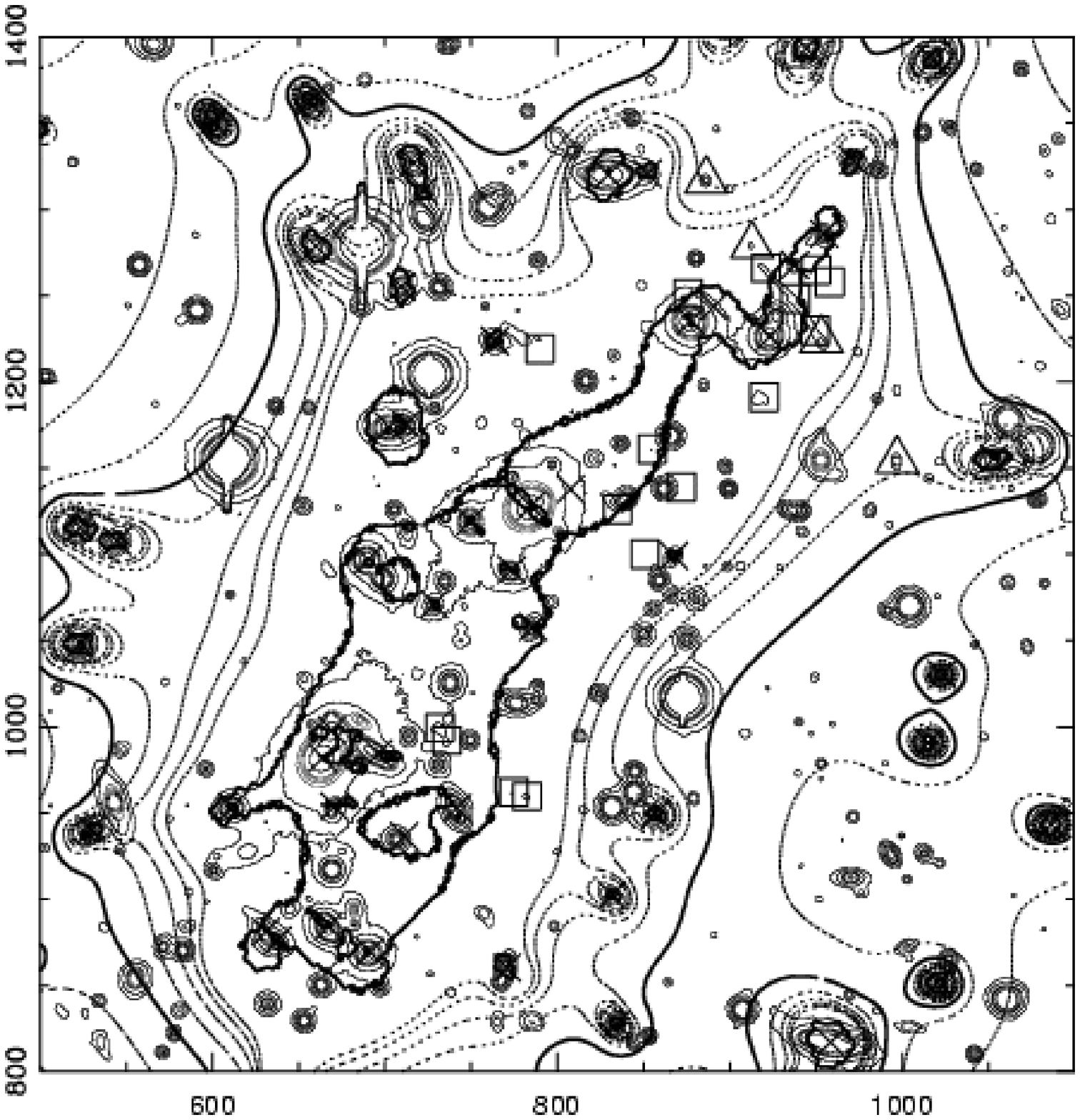} 
\hfill\parbox[b]{4.5cm}{ 
\caption{ 
The magnification contours of a source at the redshift of the arc as a 
function of the position of the image on the CCD. The contours 
% 
%%%RPS in the right-top, left-top and left bottom corners there are two 
%%%RPS dotted contours outside the thick contour 6 line. are they at level 
%%%RPS 4 and 5 or both at 5? 
% 
of magnification factors from 4 to 10 are shown; the contour of a 
magnification factor of 6 (corresponding to about 2 magnitudes) is a 
thick solid line, whereas the remaining ones are dashed lines. The 
central thick and solid contours denote the critical lines (with 
formally infinite magnification). The thin contours denote the 
isophotes of the visible objects. The squares denote the position 
of multiple images used to obtain the best fitting mass model. 
Note, that not most of the multiple images are not visible themselves on
this contour plot, since displaying these low surface brightness contours
the picture would have become too crowded.
The position of galaxies included in the models are shown as crosses. 
The triangles are on the positions of the arc and the high redshift 
galaxies A, B, C discussed in this paper. Object~J does not lie inside 
the zoomed area shown in this figure. 
\label{fig:mag_map} 
}} 
\end{figure*}

The procedure of the mass modelling was the following: 
we started from the minimum mass, that the cluster has to have, due to 
its visible member galaxies: 
about 150 early type cluster galaxies have 
been identified by their colors and then their masses have been modeled 
with 
singular isothermal spheres using the Faber-Jackson relation (Faber 
\& Jackson~\cite{FJ76}). Although the FJ-relation has a quite large 
scatter, and therefore the light deflection is not accurately described 
for 
%%%RPS individual --> individual 
individual galaxies, 
%%%RPS , the --> that 
this procedure ensures that on average the masses 
of the galaxies are taken into account correctly, and in particular 
the mean magnification caused by them is taken into account. 
The galaxy masses alone are not sufficient to 
produce the observed multiple images, however indicate the same 
substructure as a `guess by the eye' lens modelling would derive for the 
dark matter distribution. Next we admitted three additional dark 
matter distributions (modelled by EISC) at the southern and northern 
subclump of the center of E0657, and at the brighter one of the  two 
galaxies west of the arc. 
The positions of these dark matter distributions were allowed to 
vary as well, however did not change much from the optically guessed 
center. 
Next to critical lines (i.e. where multiple images and arcs arise) 
the exact mass distribution of cluster galaxies 
plays a critical role for the light deflection and magnification of 
the  sources. Therefore, we have modelled the nearest galaxies to the arc 
(the bright one west of the arc and the faint one north of it) 
by EISCs. Since we have about 16 multiple images (including the arc) 
% stella - including the arc?!, ja, 5.juli 2001, see above 
positions for the 
lens modelling, the number of free parameters was never exceeded by 
the number of constraints. The positions of the multiple images are 
marked as 
%%%RPS deleted ``a'' 
empty squares in Fig.~\ref{fig:mag_map}, and the positions 
of the masses (dark matter and cluster galaxies) are marked as crosses 
in the same Figure. 

The magnification map (for a $z=3.24$ source) 
for the best fitting model is shown in Fig.~\ref{fig:mag_map}. 
In the major part 
of the FORS field the cluster magnifies high-redshift sources by more 
than a factor of 2 in flux, in most of the central part 
(which is shown in this figure) 
the magnification is larger than two magnitudes! 

The magnification intervals of the objects in Table 1 have 
been derived by exploiting the variety of `in principle-possible' mass 
distributions. For example, instead of modelling the cluster galaxies 
by SIS, we have modelled them by elliptical isothermal spheres (with 
no core radius) and inferred the orientation and ellipticity from the 
assumption that the dark mass traces the galaxy light. Also, we have 
investigated the change of magnification if the bright 
galaxy 9\arcsec  west of source A would be a cluster member (which it 
formally is not, due to its color). In this case (as expected) only the 
magnification of A is changed, from 1.9-2.0 to 2.1-2.2 magnitudes. 
%%%RPS magnication --> magnification 
The magnification of the sources A,B,C and core are constrained by the 
lens modelling, whereas the object J is fairly away from the multiple 
image region, and there the magnification is derived from assuming 
that the dark matter model (constrained in the center of the 
cluster) is a valid description also in the more outer region of the 
cluster.

\subsection{Spectral Characteristics} 
\label{results_spec} 

The spectra of Fig.~\ref{fig:spec_arc} to \ref{fig:spec_j} are not 
unusual for high-redshift galaxies (Trager et al.~\cite{TFDO97}). 
Several lines of highly ionized metals as well as Ly$\alpha$ 
and Ly$\beta$ are seen in absorption; object~C additionally shows 
strong Ly$\alpha$ emission and object~J a weak Ly$\alpha$ emission 
component. The absence of a strong \NV\ resonance line indicates 
a temperature of the dominating stellar population of 30000\,K or 
below; furthermore, the absence of P Cygni emission at this effective 
temperature shows that supergiants do not contribute significantly to 
the integrated spectrum (Walborn, Parker, \& Nichols 1995) or that the 
metallicity might be extremely low.  The shape of the continuum redward 
of the Ly$\alpha$ line is compatible with the above characteristics; 
however, determination of the temperature using the continuum slope 
strongly depends on the assumptions made concerning the presence of 
dust and the related UV extinction. For example, assuming
a LMC~law 
for the intrinsic reddening (see Howarth 1983), 
an uncertainty of $\pm 0.05$ in $E(B-V)$ 
amounts to a change in the continuum slope that roughly corresponds to a 
temperature variation of about 5000\,K. A more conclusive 
analysis of the stellar content using spectral features will require 
re-observation of these objects with a higher S/N ratio, 
and the use of a realistic model for the stellar population.

Nevertheless, we have undertaken a first tentative step towards a 
more quantitative analysis of the observed spectra of objects~J and~A, 
where the S/N is largest and the observed spectral range redward 
of Ly$\alpha$ is large enough to determine the slope of 
the continuum flux reliably (which is not true for object~C and the arc/core). 
Based on the above arguments we have used the Munich hot star wind 
code {\em WMbasic} (Pauldrach et al.~\cite{Petal98}, \cite{Petal01}) 
to calculate the spectrum of a single 25000~K B-main-sequence star, 
adopting a typical mass-loss rate of $10^{-9} M_{\sun}/{\rm yr}$.
We have tested models with solar, half solar, and 1/5th solar
metallicity and find that the 1/5th solar metallicity model
(corresponding to the metallicity of the SMC, see Haser et al.~1998) 
gives the overall best agreement with the strengths of the observed
photospheric line features, but with the given S/N ratio we 
admit that this value 
%%%RPS is --> of 
of metallicity is not very well constrained.

\begin{figure*} 
\includegraphics[width=\textwidth]{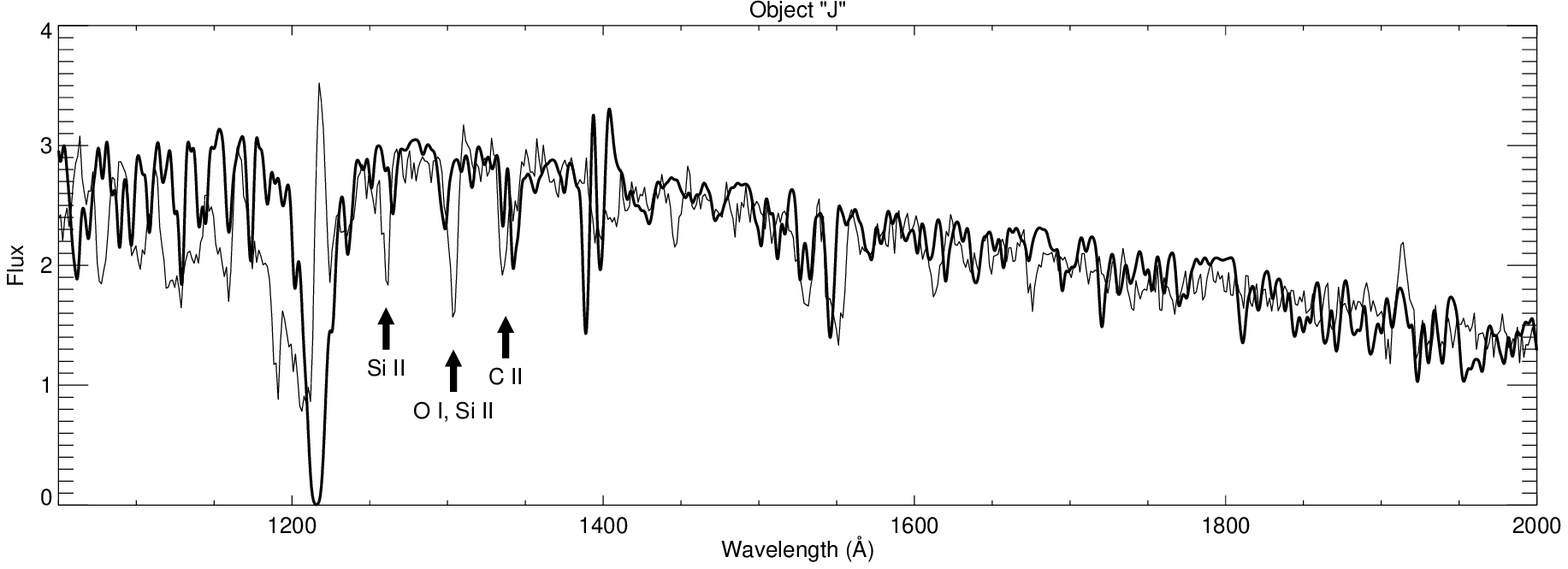} 
\includegraphics[width=\textwidth]{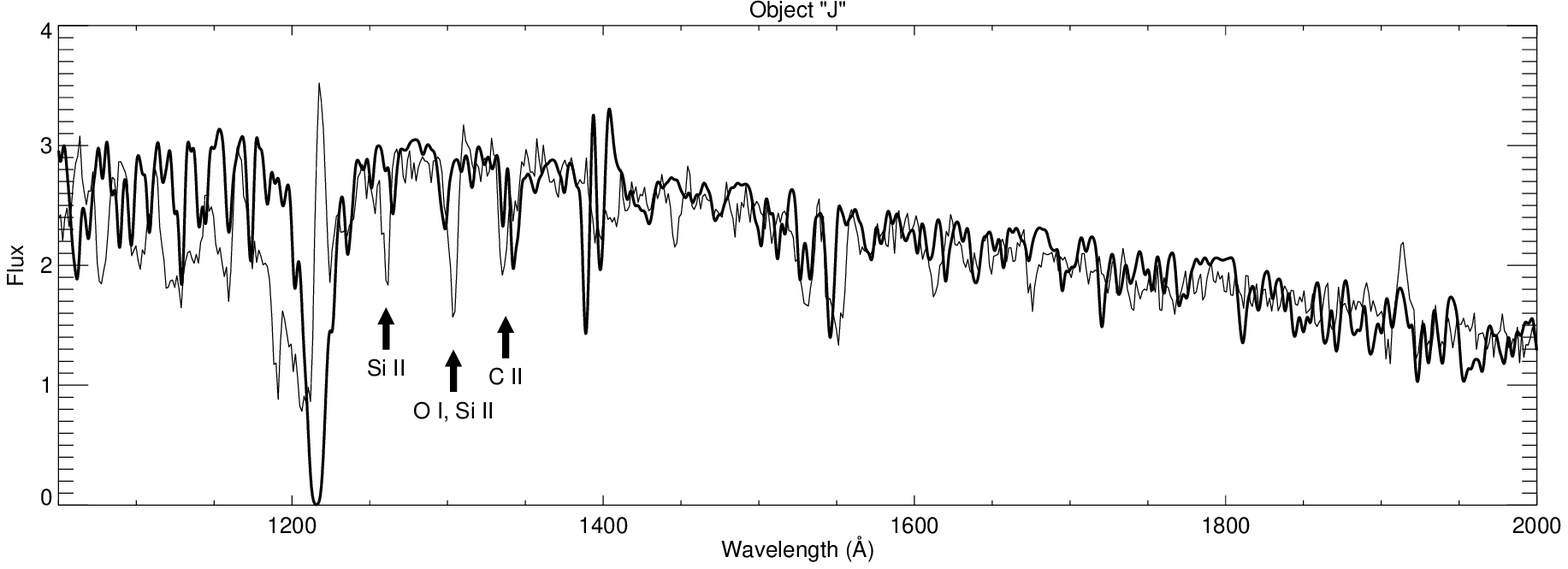} 
\caption{ 
Comparison of the observed spectra (thin lines) of objects~J at 
$z = 2.61$ (upper panel) and~A at $z = 2.34$ (lower panel) with the 
synthetic spectrum (thick line) of a single 25000~K B-main-sequence 
star adopting a mass-loss rate of $10^{-9} M_{\sun}/{\rm yr}$. 
The observed spectra have been corrected for galactic extinction with 
$E(B-V) = 0.058$ prior to blueshifting; we have additionally reddend 
the synthetic spectrum by $E(B-V) = 0.13$ (upper panel) and $0.14$ 
(lower panel) using the LMC~law (see detailed description in the text). 
Interstellar absorption of Ly${\alpha}$ has been included in the 
synthetic spectra, but not that of the other prominent interstellar 
lines (\OI, \SiII\ and \CII, indicated by arrows). 
\label{fig:calc} 
} 
\end{figure*}

Fig.~\ref{fig:calc} shows this spectrum overlaid on the spectra 
of objects~J and~A.  To account for both galactic and intrinsic 
reddening in this comparison, we have de reddened the observed spectra 
(in the optical spectral range where they were observed) using 
Howarth's~(\cite{H83}) analytical expression for Galactic extinction 
with a value of $E(B-V) = 0.058$ (Burstein and Heiles~\cite{BH84}) 
prior to blueshifting, and reddened the synthetic spectrum in its UV 
rest frame using the LMC extinction law (to allow for the expected 
lower metallicity in the observed galaxies) with $E(B-V) = 0.13$ and 
$0.14$ for objects~J and~A, respectively (the latter values are, of 
course, fit parameters).  Apart from hydrogen (with a column density 
of $10^{21}/{\rm cm}^{2}$ at the rest frame of the stellar spectrum) 
we have not yet added interstellar absorption lines to the synthetic 
spectrum, hence the obvious discrepancy to the observed spectra at 
these line positions (marked with arrows in Fig.~\ref{fig:calc}). 
Since the galaxy spectra represent the total stellar population we also 
expect the fit to improve for a superposition of synthetic spectra of 
different spectral types corresponding to a more realistic population. 
%%%RPS subject first! 
We attribute the large spectral difference between stellar model and 
galaxy spectrum 
blueward from Ly$\alpha$  to the Lyman forest. 

\subsection{Equivalent Widths} 
\label{results_ew} 

\begin{figure*} 
\includegraphics[width=12.0cm]{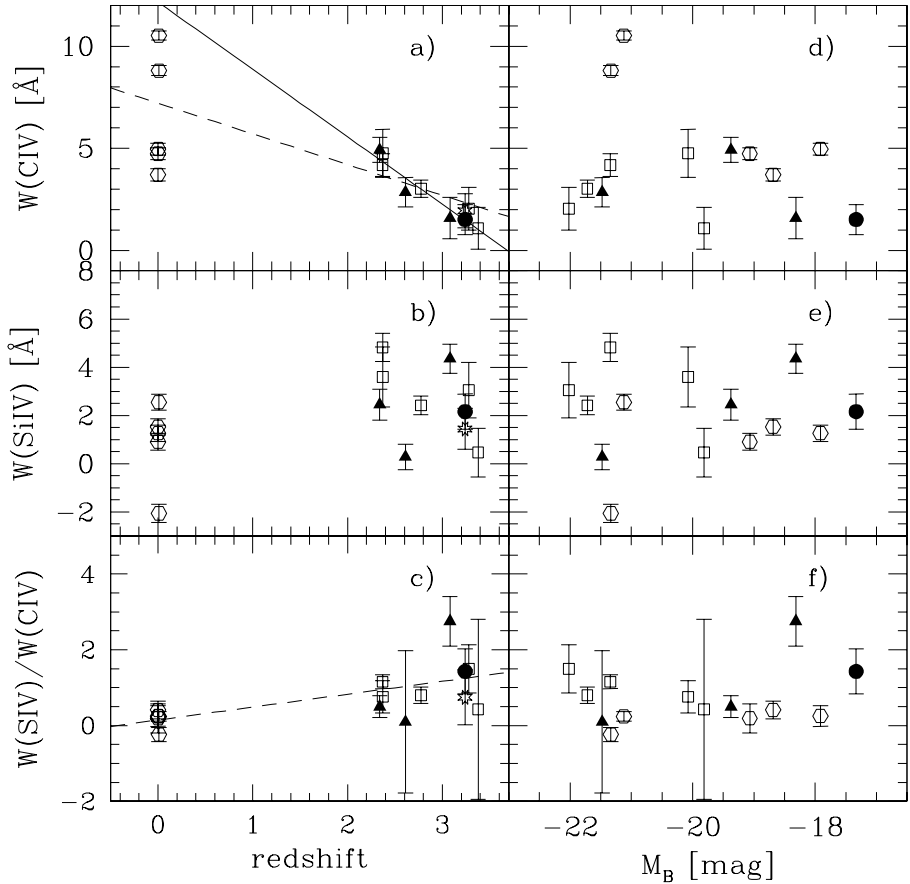} 
\hfill\parbox[b]{5.5cm}{ 
\caption{ 
Measured \CIV~1550 and \SiIV~1400 rest frame equivalent widths as well 
as their ratio \SiIV/\CIV\ versus redshift (a, b, c) and absolute 
B-magnitude (d, e, f). The absolute magnitudes are determined with 
$H_0 = 50\,\rm km/s/Mpc$ and $q_0=0.5$. For the $k$-correction 
we used values provided by M\"oller, Fritze-v.\,Alvensleben, 
and Fricke~(2001). Evolutionary corrections are not applied. 
Filled triangles: 1E0657-56 galaxies; filled circle: core; asterisk: 
arc of 1E0657-56; open squares: FDF galaxies; open hexagons: nearby 
starburst galaxies.  Solid line: weighted $\chi^2$ fit for high-$z$ 
objects only (i.e., without local ones); dashed line: weighted $\chi^2$ 
fit for all shown galaxies (both high-$z$ and local). 
\label{fig:eqwidth} 
}} 
\end{figure*} 

To investigate the stellar population of the observed galaxies 
in more detail we also measured the rest frame equivalent widths 
$W_0$ of \CIV~1550 and \SiIV~1400 from continuum-normalized 
spectra in the following way: 
\begin{equation} 
W_0 = \left( 1-\frac{S(\lambda_0)}{C(\lambda_0)} \right)d\lambda 
\end{equation} 
where 
\begin{equation} 
S(\lambda_0) =  \frac{\delta\lambda}{(1+z)\,d\lambda} 
\sum_{\lambda_i = \lambda_0-d\lambda/2}^{\lambda_0+d\lambda/2} S(\lambda_i) 
\end{equation} 
is the line flux at the central rest frame wavelength $\lambda_0$. 
For the width of the line window we chose $d\lambda = 30$\,\AA; 
$\delta \lambda = 5$\,\AA\ is the spectral pixel scale. 
The continuum flux in the continuum-normalized spectra at 
$\lambda_0$ is 
\begin{equation} 
C(\lambda_0) = 1. 
\end{equation} 
The error of the measured equivalent width is dominated by the S/N of 
the spectra (the mean error of the continuum fit is $\leq 10 \%$).
For object B 
%%%RPS dropped ``the'' 
we could not measure reliably the equivalent width 
due to the low S/N of its spectrum.

As described in Section~\ref{results_phot}, in the ``arc'' of 1E0657-56 
we see a map of the outskirts of the background galaxy at $z=3.24$, 
while the ``core'' maps mainly the total galaxy.  Hence, in principle, 
a gradient of the \CIV\ and \SiIV\ line strengths could be present in 
the arc.  But as seen in Figs.~\ref{fig:eqwidth}a to~c 
%%%RPS (green symbols) --> filled circle and asterix 
(filled circle and asterix) 
there is no significant difference of the stellar population in the 
central region and the outskirts of this high-$z$ galaxy. 

To increase the sample and to compare our lensed galaxies with high-$z$ 
field galaxies we included in Fig.~\ref{fig:eqwidth} measurements 
for some FORS Deep Field (FDF, see Appenzeller et al.~\cite{Aetal00}) 
galaxies. The FDF objects were observed with the same spectroscopic 
setup as the galaxies in 1E0657-56 (see Sect.~\ref{observations}), 
but using the multi object mask instead of the long slit.  The data 
reduction was performed as described in Sect.~\ref{observations}. 
To select a subsample of high-$z$ galaxies among the FDF 
sample we used following criteria: $I \leq 24.5$\,mag and $z \geq 
2$. Additionally we needed spectra of sufficient S/N to determine 
reliable equivalent widths.

Figure~\ref{fig:eqwidth}a shows that for $z > 2$, $W(\CIV)$ increases 
with decreasing redshift.  A weighted $\chi^2$ fit gives a slope of 
$\alpha = -3.31 \pm 0.10$.  To continue this relation to $z=0$ we 
also measured the equivalent width of 5 nearby starburst galaxies 
(NGC~1705, 4228, 5253, 7673, 7714) arbitrarily chosen from IUE 
archive\footnote{\tt http://ines.vilspa.esa.es}. The spectra were 
obtained by the Short Wavelength Prime Camera (SWP; $\lambda = 
1150\dots1980$\,\AA) 
% 
%%%RPS the resolution of the high redshift spectra is a factor 3-4 higher. 
%%%RPS does this affect the equivalent widths? (only 5 pixels available 
%%%RPS in the line window at z=0). 
% 
in the low dispersion mode (6\,\AA) and with 
the large aperture ($10'' \times 20''$). The standard reductions of the 
data was done via the pipeline provided by the archive. The equivalent 
widths were measures as described above.  If the corresponding 
data points at $z=0$ are added to Fig.~\ref{fig:eqwidth}, the trend 
persists although the linear correlation is less pronounced ($\alpha = 
-1.50 \pm 0.10$).  However, if correct, the relation between the \CIV\ 
strength with $z$ is expected not to be linear anyway.  Interestingly, 
the scatter of the \CIV\ equivalent width at $z=0$ is much larger 
than at high $z$.  This apparent correlation is not caused by the 
galaxies' luminosity, since the absolute magnitude of our galaxies do 
not show any significant trend with their redshift.  Since $W(\CIV)$ 
mainly depends on the metallicity of the stellar population but only 
little on the age of a starburst (Leitherer et al.~\cite{Letal01}) 
the correlation with $z$ could indicate an increase of metallicity 
with decreasing redshift (i.e., increasing age of the universe) 
in the observed high luminosity galaxies.  In this respect these 
galaxies seem to behave similar to the damped Ly${\alpha}$ systems 
(Savaglio et al.~\cite{Saetal01}).  The fact that the scatter for 
\SiIV\ is much larger at all redshifts and that no correlation with $z$ 
is present for this line may have 
%%%RPS added ``the'' 
the following reasons: While $W(\CIV)$ 
is universally present for O~supergiants, main sequence stars and 
dwarfs, $W(\SiIV)$ is luminosity dependent and rapidly decreases from 
supergiants to dwarfs (Walborn \& Panek~1984; Pauldrach et al.~1990). 
Therefore the \SiIV\ is more strongly affected by population 
differences (i.e., stellar age differences) than the \CIV\ line. 
$W(\SiIV)$ has its maximum in B0/B1 stars, while $W(\CIV)$ is mainly 
universally present in O and bright B~stars (e.g., Leitherer, Robert, 
and Heckman~\cite{LRH95}).  Hence the ratio of $W(\SiIV)/W(\CIV)$ 
contains information about the star formation history (instantaneous 
or continuous) and the stellar population itself (e.g., IMF and cutoff 
masses).  Unfortunately the different parameters that determine the 
value of $W(\SiIV)/W(\CIV)$ cannot be disentangled easily. 

As seen in Fig.~\ref{fig:eqwidth}c the ratios of $W(\SiIV)/W(\CIV)$ for 
our high-$z$ galaxies are randomly scattered and show no detectable 
trend with redshift.  However, if the local starburst galaxies are 
included a trend corresponding to increasing ratio with $z$ seems to 
be present (a weighted $\chi^2$ fit gives $\alpha = 0.34 \pm 0.06$). 
This could be understood in terms of increase of the relative 
importance of continuous star formation (decrease of instantaneous 
starbursts) at low $z$.  To test the presence of this correlation we 
have started to investigate a larger sample of high-$z$ galaxies from 
our FDF survey ($z > 1$). 

Finally, Figs.~\ref{fig:eqwidth}d to f show that there is no overall 
dependence of the measured equivalent width on the galaxies' 
luminosity. If in fact $W(\CIV)$ is mainly determined by the 
metallicity, the nearby starburst galaxies seem to follow the well 
known local metallicity-luminosity relation (e.g. Kobulnicky \& 
Zaritsky 1998), while the high-$z$ galaxies do not conform to this 
relation. As also found for Lyman break galaxies by Pettini et al. 
(2001) the high-$z$ galaxies investigated in this paper seem to be 
overluminous for their metallicity ($W(\CIV)$), which may indicate that 
their mass-to-light ratios are low compared to present-day galaxies. 
Figs.~\ref{fig:eqwidth}d to f also show that our four galaxies 
magnified by 1E0657-56 belong to the intrinsically faintest starburst 
galaxies at high redshift that have been investigated spectroscopically 
so far.  In this respect the present investigation provides a good 
example for the scientific potential of ``gravitational telescopes'' 
for the study of distant galaxies. 

\section{Conclusions} 
\label{conclusions} 

We report on the discovery of five high redshift ($z > 2$) galaxies 
in the background of the cluster 1E0657-56 which are gravitationally 
lensed and thus magnified and amplified by the cluster.  The most 
interesting object in the field is the prominent arc+core object 
lying NW of the clusters center.  The core represents the main 
part of a high redshift background galaxy, while some part of the 
galaxy's outer region is mapped into three merging images (the arc). 
We briefly discuss the lens model which yields a flux amplification 
of the background sources by a factor up to 20 in flux.

The spectra we obtained for the arc+core component and for four 
other objects confirm that these background sources are indeed 
young (starburst) galaxies at high redshift.  Besides Ly$\alpha$ the 
absorption features of several highly ionized metals are identified. 
Two objects show additional prominent Ly$\alpha$ emission. 
A comparison with calculated stellar spectra shows that the light 
of these galaxies is dominated by hot ($\approx 25000\,\rm K$) 
low metallicity ($\approx$ 1/5 solar) stars. 

%%%RPS rephrased to clarify the object sample studied 
The measurements of the \CIV\ and \SiIV\ equivalent widths 
of the objects studied here plus a sample of nearby and high z galaxies 
show a strong correlation of \CIV\ with redshift.  Though rather 
speculative at this stage, we interpret this as a result of the 
increase of metallicity with the age of the universe.  The absence of 
any correlation of \SiIV\ with redshift as well as a little evolution 
of the \SiIV/\CIV\ ratio points to variations in the star formation 
history and/or an evolution of the stellar population with redshift. 
These results will have to be confirmed by a larger sample of high-$z$ 
galaxies. 

\begin{acknowledgements} 
We thank E.~Falco and M.~Ramella for pointing out this cluster 
to us as a possible target for our project and also for 
providing us with their previous NTT images. 
This research was supported by the 
Sonderforschungsbereiche 375 and 439, 
by the DFG in the ``Gerhard-Hess-Programm'' under grant Pa~477/2-3, 
and by the DLR under grant 50~QV~9704~1. 
Some image reduction was done using the MIDAS 
and/or IRAF/STSDAS packages. 
\end{acknowledgements}

\end{document}